# QUALINET
# White Paper on Definitions of Immersive Media Experience (IMEx)

**List of Authors and Contributors**
Andrew Perkis (andrew.perkis@ntnu.no, editor)
Christian Timmerer (christian.timmerer@itec.uni-klu.ac.at, editor), Sabina Baraković, Jasmina Baraković Husić, Søren Bech, Sebastian Bosse, Jean Botev, Kjell Brunnström, Luis Cruz, Katrien De Moor, Andrea de Polo Saibanti, Wouter Durnez, Sebastian Egger-Lampl, Ulrich Engelke, Tiago H. Falk, Jesús Gutiérrez, Asim Hameed, Andrew Hines, Tanja Kojic, Dragan Kukolj, Eirini Liotou, Dragorad Milovanovic, Sebastian Möller, Niall Murray, Babak Naderi, Manuela Pereira, Stuart Perry, Antonio Pinheiro, Andres Pinilla, Alexander Raake, Sarvesh Rajesh Agrawal, Ulrich Reiter, Rafael Rodrigues, Raimund Schatz, Peter Schelkens, Steven Schmidt, Saeed Shafiee Sabet, Ashutosh Singla, Lea Skorin-Kapov, Mirko Suznjevic, Stefan Uhrig, Sara Vlahović, Jan-Niklas Voigt-Antons, Saman Zadtootaghaj.

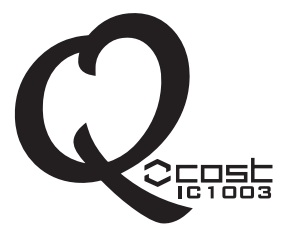

Qualinet 2020

# Table of Contents

This White Paper is a contribution by QUALINET, the European Network on Quality of Experience in Multimedia Systems and Services (http://www.qualinet.eu/) to the discussions related to Immersive Media Experience (IMEx). It is motivated by the need for definitions around this term to foster a deeper understanding of ideas and concepts originating from multidisciplinary groups but with a joint interest in multimedia experiences. Thus, this white paper has been created mainly with such multimedia experiences in mind but may be also used beyond.

The QUALINET community aims at extending the notion of network-centric Quality of Service (QoS) in multimedia systems, by relying on the concept of Quality of Experience (QoE). The main scientific objective is the development of methodologies for subjective and objective quality metrics taking into account current and new trends in multimedia communication systems as witnessed by the appearance of new types of content and interactions. QUALINET (2010-2014 as COST Action IC1003) meets once a year collocated with QoMEX (http://qomex.org/) to coordinate its activities around 4 Working Groups (WGs): (i) research, (ii) standardization, (iii) training, and (iv) innovation.

The white paper has been created based on an activity launched at the 13th QUALINET meeting on June 4, 2019 in Berlin (collocated with QoMEX2019) with Andrew Perkis and Christian Timmerer appointed as editors and hosted as part of the Task Force 7, Immersive Media Experiences (IMEx). The editors created a draft table of contents followed by an invitation for contributions where 10 section leads have been appointed to coordinate inputs from 44 contributing authors. On December 10, 2019, a consolidated first draft has been released among all section leads and editors for internal review.

After incorporating the feedback from all section leads, the editors released a version on January 22, 2020 for QUALINET community review. After receiving feedback from QUALINET at large and incorporating it, the editors distributed the white paper widely for an open, public community review on February 22, 2020 (e.g., research communities/committees in ACM and IEEE, standards development organizations, various open email reflectors related to this topic). The feedback received from this public consultation process resulted in the final version which has been approved during the 14th QUALINET meeting on May 25, 2020 in a virtual/online meeting (collocated with QOMEX2020).

# Introduction

**Section Lead**
Andrew Perkis &
Christian Timmerer

With the coming of age of virtual/augmented reality and interactive media, numerous definitions, frameworks, and models of immersion have emerged across different fields ranging from computer graphics to literary works. Immersion is oftentimes used interchangeably with presence as both concepts are closely related. However, there are noticeable interdisciplinary differences regarding definitions, scope, and constituents that are required to be addressed so that a coherent understanding of the concepts can be achieved. Such consensus is vital for paving the directionality of the future of immersive media experiences (IMEx) and all related matters.

The aim of this white paper is to provide survey of definitions of immersion and presence which leads to a definition of immersive media experience (IMEx). The Quality of Experience (QoE) for immersive media is described by establishing a relationship between the concepts of QoE and IMEx followed by application areas of immersive media experience. Influencing factors on immersive media experience are elaborated as well as the assessment of immersive media experience. Finally, standardization activities related to IMEx are highlighted and the white paper is concluded with an outlook related to future developments.

# Survey of Definitions of Immersion and Presence

**Section Lead**
Katrien De Moor, Sebastian Egger-Lampl & Alex Raake

**Authors**
Søren Bech, Katrien De Moor Wouter Durnez, Sebastian Egger-Lampl, Babak Naderi, Alex Raake, Sarvesh Rajesh Agrawal, & Steven Schmidt

Taken literally, to immerse means to plunge into something that surrounds or covers (refer to Merriam-Webster). As far as its metaphoric use in academic literature is concerned, this definition is as far as the consensus goes. In the past decades, various domains have lent their own interpretations to the concept. From a bird's eye view, two broad perspectives can be distinguished. The first perspective classifies immersion as a system property, as is exemplified in the following definition[1]:

> [Immersion] refers to the degree to which immersive media environments sub-merges the perceptual system of the user in computer-generated stimuli. The more the system blocks out stimuli from the physical world, the more the system is considered to be immersive.

This viewpoint, which can be traced back to initial research on telepresence equates immersion to the system's ability to provide a user's senses with surrogate stimuli replacing or complementing real-life signal input. To do so, immersive systems may use technology such as displays (e.g., VR, AR, 4k HDR, 8k etc.), accurate positional tracking, and haptic feedback. In another work by van Gisbergen[2], immersion is created through six P dimensions, namely presence, perspective, proximity, point of view, participation, and place.

From a second perspective, immersion is defined as the user's response – cognitive or otherwise – to specific characteristics of the system or content. Nilsson[3] outlines three further subcategories. The first is referred to as perceptual or sensory immersion. It is used when describing a response to the technical characteristics of a virtual. The remaining views on immersion dissociate the term from any such technological substrate, focusing on content characteristics instead. The term fictional or narrative immersion is used to describe a user's response to narrative elements, such as the story, the world in which it unfolds, or the characters it features. Challenge-based immersion, finally, is similar to the concept of flow, as it reflects a state of mental absorption in response to challenges that match the user's skill level.

Presence (its conceptualization, measurement, determinants, and influence factors) has become a prominent topic in several research domains, e.g., in research on digital games, virtual and augmented reality, and computer-supported collaborative work over the last few decades. However, as different scientific communities use distinct terms (e.g., presence, telepresence, mediated presence, virtual presence), there is a lack of a unified terminology and a range of definitions and theoretical models of presence. While presence tends to be confused with immersion and the two have even been considered synonymous, we try to make a distinction between the terms. This overview of presence is limited to the most prominent conceptualizations, but a thorough overview can be found in Lee (2006)[4].

The most dominant definitions of presence associate it with a "sense of being there" (in the virtual environment)"[5] and with the suspension of disbelief, as used by

1   Biocca, F., & Delaney, B. (1995). Immersive virtual reality technology. Communication in the age of virtual reality, 15, 32.

2   Van Gisbergen, M. S. (2016). Contextual connected media: How rearranging a media puzzle, brings virtual reality into being.

3   Nilsson, N. C., Nordahl, R., & Serafin, S. (2016). Immersion revisited: a review of existing definitions of immersion and their relation to different theories of presence. Human Technology, 12(2).

4   Lee, K. M. (2004). Presence, explicated. Communication theory, 14(1), 27-50.

5   Slater, M., & Usoh, M. (1993). Representations systems, perceptual position, and presence in immersive virtual environments. Presence: Tele-operators & Virtual Environments, 2(3), 221-233.

Slater et al.[6] when referring to "place illusion". Witmer and Singer[7] further specified presence as "the subjective experience of being in one place or environment, even when one is physically situated in another"[8]. Slater[9] instead refers to the "illusion" of presence and uses "place illusion" to depict the above type of presence. It refers to a very strong illusion of being in a place, even though knowing very surely that one is not really there (which may contribute to a willingness to suspend disbelief). Lombard and Ditton[10] performed a broad literature review resulting in six different conceptualizations of presence. They broadly defined presence as "the perceptual illusion of non-mediation" in an attempt to encompass the various conceptualizations of presence in the literature in one definition.

The most prominent classifications of types of presence, according to Biocca, refer to physical presence (i.e., the sense of physically being located in another place), social presence (i.e., the sense of being together with a virtual or remotely located communication partner), and at their intersection: co-presence ("a sense of being together in a shared space at the same time"). Other conceptualizations similarly distinguish between physical and social presence, but also introduced the notion of self-presence (i.e., "the awareness of self-identity inside a virtual world". However, as also argued by Lombard and colleagues, there is still a lack of understanding of how different types/dimensions of presence interplay with each other and what this implies for the overall experience of presence.

---

6  Slater, M. (2009). Place illusion and plausibility can lead to realistic behaviour in immersive virtual environments. Philosophical Transactions of the Royal Society B: Biological Sciences, 364(1535), 3549-3557.

7  Witmer, B. G., & Singer, M. J. (1998). Measuring presence in virtual environments: A presence questionnaire. Presence, 7(3), 225-240.

8  Note: The understanding of spatial presence in these definitions presupposes (and thus implicitly contains) a sense of temporal presence.

9  Slater, M. (2018). Immersion and the illusion of presence in virtual reality. British Journal of Psychology, 109(3), 431-433.

10  Lombard, M.,& Ditton, T. (1997). At the heart of it all: the concept of presence. Journal of computer-mediated communication, 3(2), JCMC321

# Definition of Immersive Media Experience (imex)

**Section Lead**
Asim Hameed

**Authors**
Jean Botev, Kjell Brunnström, Asim Hameed, Tanja Kojic, & Antonio Pinheiro

Immersive media have drawn considerable interdisciplinary interest over the past few decades, which successfully delivered various prolific frameworks for immersive media[11]. They involve multi-modal human-computer interaction where either a user is immersed inside a digital/virtual space or digital/virtual artefacts become a part of the physical world. Immersive media invoke a user's sense of being there (i.e., presence according to the definition above). From an experiential perspective, this combines the physical and psychological concepts of immersion, immediacy, and presence that are foundational to fully comprehend an immersive media experience[12].

Different frameworks draw a consensus on features characterizing immersive media, these are[13]:

1. *Immersivity,* the combination of sensory (physical/system) cues with symbolic (content) cues essential for user emplacement and engagement.

2. *Interactivity,* with digital/virtual artefacts and avatars through an interface.

3. *Explorability,* the possibility for users to move freely and discover the world offered.

4. *Believability,* the fidelity and validity of sensory features within the generated environments, e.g., photorealism.

5. *Plausibility,* within generated environments is the coherence and consistency of symbolic features (ideas, relationships, etc.) for the user to form mental concepts.

In continuation, we summarize immersive media as:

**a high-fidelity simulation provided and communicated to the user through multiple sensory and semiotic modalities. Users are emplaced in a technology-driven environment with the possibility to actively partake and participate in the information and experiences dispensed by the generated world.**

The development of dedicated capturing technologies, processing and delivery mechanism has always been instrumental in creating and advancing IMEx. Immersive Media Technologies (IMT) attempt to emulate a real-world through a digital or simulated recreation, resulting in a spatial illusion or sense of presence. Originating in the mid-20th century, IMT caught resurgent public and research attention in the 1990s and – due to the rapid advances and miniaturization in mobile hardware – during the past decade, finding broader implementation and application.

IMT can be located at different positions along the so-called *virtuality continuum*[14], with experience contexts ranging from real environments over mixed reality approaches to fully virtual environments. Currently, IMT manifest themselves either as providing non-interactive spherical content for multi-directional viewing (e.g., 360 or 360-VR) and as interactive extended reality (XR), including:

---

11   Steuer, J. (1992). Defining virtual reality: Dimensions determining telepresence. Journal of communication, 42(4), 73-93.

12   Witmer, B. G., & Singer, M. J. (1998). Measuring presence in virtual environments: A presence questionnaire. Presence, 7(3), 225-240.

13   Schuemie, M. J., Van Der Straaten, P., Krijn, M., & Van Der Mast, C. A. (2001). Research on presence in virtual reality: A survey. CyberPsychology & Behavior, 4(2), 183-201.

14   Milgram, P., Takemura, H., Utsumi, A., & Kishino, F. (1995, December). Augmented reality: A class of displays on the reality-virtuality continuum. In Telemanipulator and telepresence technologies (Vol. 2351, pp. 282-292). International Society for Optics and Photonics.

1. **Mixed Reality (MR)** combines real and virtual content, registered in various virtual spaces, including 2D and 4D, allowing for real-time interaction; according to perspective
    a. Augmented Reality (AR) overlays computer-generated content on the real space;
    a. Augmented Virtuality (AV) incorporates real objects into a virtual space;
2. **Virtual Reality (VR)** occludes physical space to provide interactive and non-interactive experiences of a fully computer-simulated "virtual" world or a photographically "captured" real world.

Note that the above definitions are independent of hardware type and not solely limited to visual media. Also, IMT can be further classified into subcategories according to the targeted degree of immersion. For instance, the goal of VR generally is full immersion, while IMT in an AR context focuses more on the provision of engaging, participatory experiences embedded in the real space.

Immersive media are widely understood from an experiential perspective as a user's sense of presence achieved through various types of immersion (technical, challenge-based, etc.). This potential can thus be achieved through an interwoven triad of immersivity, interactivity, and narrativity.

# Quality of Experience (QoE) for Immersive Media

**Section Lead**
Sebastian Möller

**Authors**
Sabina Baraković, Jasmina Baraković Husić, Kjell Brunnström, Tiago H. Falk, Sebastian Möller, Antonio Pinheiro, Alexander Raake, Rafael Rodrigues, & Stefan Uhrig

In this section, we will establish a relationship between the concepts of Quality of Experience (QoE) and immersive media experience (IMEx). In order to understand how QoE is formed by a human user, the framework from the Qualinet White Paper[15] provides useful guidance. It describes experienced quality as a result of a comparison and judgment process, in which the perceived quality features, resulting from a perception and reflection process triggered by a physical signal, are compared to the desired quality features underlying the user's expectations. Along these lines, we define Quality of Experience (QoE) for immersive media as "the degree of delight or annoyance of the user of an application or service which involves an immersive media experience. It results from the fulfillment of his or her expectations with respect to the utility and/or enjoyment of the application or service in the light of the user's personality and current state."

With respect to immersive media, there are some particularities, mainly with respect to the experiences themselves as well as the influencing factors of the system, the human user, and the context of use. Immersive media often make use of virtual or augmented realities to create or augment a virtual context of use. As a consequence, a clear separation of context and system influencing factors is not possible. Instead, there is an overlay between the virtual and real context, one of which might be dominant (depending on the position of the IMT on the virtuality continuum). In addition, the service might react and adapt depending on the physical context (e.g., a location-based augmented reality service) or the social context (e.g., by connecting other players in a gaming service).

Human influencing factors, which are particularly important for immersive media, include mostly perceptual characteristics such as visual and auditory acuity. Potential impairments may reduce the perception in one perceptual modality, and in case they occur asymmetrically, an impact on the spatial perception may occur. Incongruence and delay between perceptual modalities may lead to a loss of spatial awareness, and may result in dizziness or nausea, which will strongly reduce a positive experience. These uncomfortable symptoms are usually referred to as "cybersickness", and vary between users, meaning that some are more sensitive and easily get sick while experiencing movements, while others may not. Simulator sickness depends on stimulus duration as well as on the user's prior exposure to virtual or augmented environments, beyond technical system, content and possibly context factors. The tendency to become immersed varies between users, thus can be considered as an important human influence factor as well.

With respect to IMEx, it is commonly assumed that a sense of presence is crucial for QoE of immersive media. The "sense of being there", understood as self-location within the virtual reality, is normally accompanied by a sense of virtual embodiment, which includes taking ownership of one's own virtual body (given that a body has been implemented in the virtual environment) and feeling agency when moving inside it and interacting with the surrounding world. The "place illusion" lets the user perceives these virtual objects as if they were really existing and, as a consequence, adjusts his/her behavior towards them like he/she would do towards objects in the physical world. Finally, the interaction experience between a user and the virtual reality can have a significant impact to the overall QoE. Good interactions can lead to a high-quality experience with a strong sense of presence. On the other hand, an unrealistic interaction may lead to discomfort and distress. Especially related to gaming QoE, the quality aspects controllability, immediate feedback, and responsiveness of a system are used to describe the input quality.

---

15    Le Callet, P., Möller, S., & Perkis, A. (2012). Qualinet white paper on definitions of quality of experience. European network on quality of experience in multimedia systems and services (COST Action IC 1003), 3(2012).

# Application Areas of Immersive Media Experience

**Section Lead**
Antonio Pinheiro

**Authors**
Luis Cruz, Manuela Pereira, Stuart Perry, Ulrich Reiter, Rafael Rodrigues, Antonio Pinheiro, Saeed Shafiee Sabet, & Saman Zadtootaghaj

The introduction of new technologies, namely VR, AR, omni-directional video and audio and the evolution of plenoptic representations (like point clouds, light fields, or digital holography) has created new opportunities for immersive applications. Head mounted displays are considered the most important devices for immersive content distribution, providing free viewing, and even the sensation of being inside the scene. In addition, immersive theaters are being explored, providing omni-directional video and audio, and other 3D representations. Furthermore, the concept of mulsemedia[16] can be explored, where new sensory effects like haptics or olfactory are added to the visual and audio information. Those are synchronized with the multimedia content, leading to new levels of sensory immersion, improving the sensation of presence and interaction.

In this section, a classification of different immersive applications is proposed based on the level of interaction and human senses as shown in Figure 1.

Humans have five basic senses – sight, hearing, smell, taste, and touch –, which form the sensory experience. Using more senses does not necessarily mean an application is more immersive or superior to another[17]. However, information can be conveyed faster using more senses, which can help in creating a sense of full immersion in a shorter period of time.

Immersive media can also be classified according to the interaction level. The level of interaction while consuming media can be anywhere between being passive, active and interactive. Passive is when the user gets different sensory experiences without making specific actions (i.e., no gesture, no movements). Active, in turn, is when the user creates sensory activities (i.e., screaming, jumping, etc.). Lastly, interactive is when a flow of sensory activities is created between the users and their surroundings.

In immersive technologies applications, interactivity can be considered as a product for human-system and system-system interaction, or as a process for human to human interaction.

There are many well-known immersive applications and associated technologies. This list is not exhaustive and is intended to provide a short description of a selection of applications.

1. *Interactive digital storytelling:* Immersive narratives are present in all application domains as a driver for the story. This could be for journalism, historians, entertainment, for health and all the way to industrial application and learning

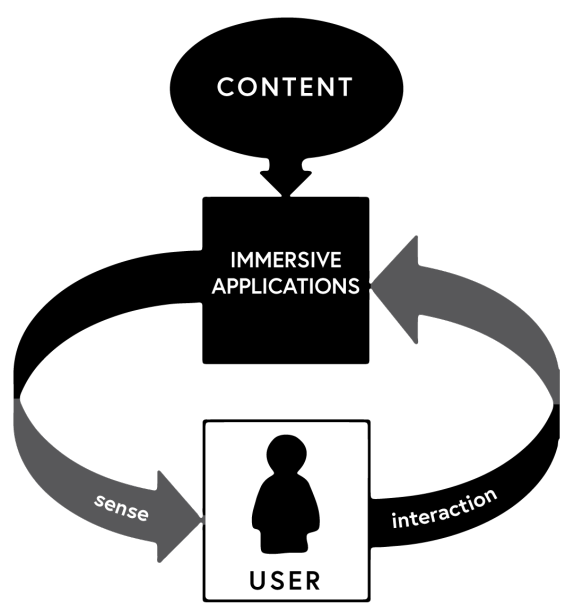

*Fig. 1 The user and the immersive application*

---

16    Ghinea, G., Timmerer, C., Lin, W., & Gulliver, S. R. (2014). Mulsemedia: State of the art, perspectives, and challenges. ACM Transactions on Multimedia Computing, Communications, and Applications (TOMM), 11(1s), 1-23.

17    Gander, P. (1999). Two myths about immersion in new storytelling media. Lund University.

2. *Gaming:* Immersion in video games gets influenced strongly by the game design process if it provides multiple channels of sensory information (sensory level), a cognitively demanding environment (level of interaction), a high level of narration, social aspects, and if universal psychological needs are fulfilled

3. *Omni-directional audio:* Omni-directional audio technologies are strongly related to the concept of object-based audio, an approach that leaves behind the restrictions of channel assignments and the corresponding production techniques dependent on a predefined number of audio tracks and related loudspeaker channels.

4. *Omni-directional video:* Omni-directional Video (OV) is defined as video footage captured across at least a 360-degree horizontal field of view.

5. *AR/VR/MR Communication/Telemeetings:* Using omni-directional audio and video for communication and collaboration purposes such as telemeetings all the way up to advanced interactive collaborative industrial design processes. Usage can span from industry to medicine.

6. *Immersive theaters:* The term "Immersive Theaters" invokes different ideas to different audiences. In the arts, this can refer to the concept of performances with strong audience participation, or simply the presentation of Omni-directional Video.

7. *Health:* Despite the difficult challenges inherent to the specificity of health applications, virtual reality and robotics have helped to shape the status quo and future perspectives in healthcare.

# Influencing Factors on Immersive Media Experience

**Section Lead**
Lea Skorin-Kapov

**Authors**
Sabina Baraković, Jasmina Baraković Husić, Kjell Brunnström, Ulrich Engelke, Eirini Liotou, Manuela Pereira, Antonio Pinheiro, Steven Schmidt, Saeed Shafiee Sabet, Lea Skorin-Kapov, Sara Vlahović, & Saman Zadtootaghaj

In this section, we describe factors influencing immersive media experiences, defined as follows: Influence Factor (IF): Any characteristic of a user, system, service, application, or context whose actual state or setting may have an influence on the immersive media experience of the user.

In line with the definition in the white paper on QoE, we consider a Human IF as any property or characteristic of a human user which influences a user's degree of (IMEx). The characteristics can be dispositional as well as variant. While some Human IFs are required for a person to become immersed in a medium, others can strengthen or weaken the experience. The fact that not every human becomes equally immersed in the same book, movie, or game, illustrates that Human IFs are of very high relevance for an IMEx.

When considering IMEx as a property of the system and perceptual response, the abilities of a user, in terms of perceptual sensitivity (e.g., visual and auditory acuity), to perceive the technology mediating the experience as well as a user's expectations towards the presentation are important. The expectations themselves can be influenced by the systems that users were confronted with in the past. Similar to QoE which is based on a comparison of perceived features with expected features, this also holds true for an IMEx and explains why in general, novel systems are often more immersive. In cases of very realistic virtual environments, e.g., due to the use of HMs), also the technical affinity of a user can play a role, which may cause a higher level of sensation for users who are very interested in technology per se. Furthermore, the sensitivity of a user towards incongruencies and timing differences between perceptual modalities (multisensory integration) is an important Human IF, which can also lead to simulator sickness and degradation of the IMEx. We also have System IFs referring to properties and characteristics that determine the technically produced quality of an application or service. They can be categorized as network-, device-, content-, and media configuration-related.

Finally, we consider Context IFs as being factors that embrace any situational property to describe the user's environment in terms of physical, temporal, social, economic, task, and technical characteristics. For a review and analysis of context influence factors related to QoE, the reader is referred to Reiter et al.[18] based on the classification of Jumisko-Pyykko et al.[19], but also considering the classification provided by Rahman et al.[20]. Different levels can be established for each of these characteristics, micro to macro for magnitude, static vs. dynamic for behavior, and rhythmic vs. random for the occurrence.

---

18    Reiter, U., Brunnström, K., De Moor, K., Larabi, M. C., Pereira, M., Pinheiro, A., ... & Zgank, A. (2014). Factors influencing quality of experience. In Quality of experience (pp. 55-72). Springer, Cham.

19    Strohmeier, D., Jumisko-Pyykkö, S., & Kunze, K. (2010). Open profiling of quality: a mixed method approach to understanding multimodal quality perception. Advances in multimedia, 2010.

20    Rahman, M. A., El Saddik, A., & Gueaieb, W. (2010). Augmenting context awareness by combining body sensor networks and social networks. IEEE Transactions on Instrumentation and Measurement, 60(2), 345-353.

# Assessment of Immersive Media Experience

**Section Lead**
Sebastian Bosse, Raimund Schatz
& Stefan Uhrig

**Authors**
Sebastian Bosse, Kjell Brunnström, Tiago H. Falk, Jesús Gutiérrez, Ulrich Engelke, Andres Pinilla Palacios, Antonio Pinheiro, Ashutosh Singla, Raimund Schatz, Steven Schmidt, Mirko Suznjevic, Stefan Uhrig, Sara Vlahovic, & Jan-Niklas Voigt-Antons

Proper examination of multimedia systems, applications and services in terms of their immersive capacity requires the availability of measurement instruments and test protocols for the assessment of different aspects of IMEx.

Three strands of assessment methods are identifiable in the literature on IMEx and related research fields: subjective, behavioral, and psycho-physiological. Each strand is based on a very different form of measurement, namely human participants' conscious introspection of their own subjective experiences, registration of behavioral responses and the physiological state of participants. Analogous to QoE assessment, defined stimuli are presented to participants who either intentionally (e.g., by answering rating scales) or non-intentionally (e.g., through behavioral actions or physiological responses) provide qualitative and/or quantitative measures of different aspects of their evoked IMEx. However, definition of stimuli and proper measurements is usually more challenging when conducting assessments under realistic conditions of IMT use. Therefore, the goals of IMEx assessment should be to develop measurement instruments and derive IMEx metrics that are ecologically valid with regard to the requirements of IMEx (e.g., in terms of interactivity), but still ensure sufficient reliability (e.g., in terms of signal-to-noise level) and diagnostic utility.

Aforementioned three strands of assessment methods for immersive media experiences can be characterized as follows:

1. *Subjective assessment* refers to the systematic elicitation and subsequent analysis of human participants' opinions on the experience of an immersive stimulus. This involves explicitly inquiring their feedback with regard to specific aspects of interest of an immersive media experience, typically reported using questionnaires, rating devices (such as sliders, dials or gloves) or structured interviews that trigger necessary introspection processes and capture qualitative and quantitative results.

2. *Behavioural assessment* is based on observing and tracking user behaviours, such as physical movement, social interaction and different in-application choices. These behaviours are evoked as a non-intentional, automatic response and do not necessitate conscious introspection like subjective methods.

3. The usage of *psycho-physiological methods* promises to provide alternative, yet direct measures of immersion and presence aspects that overcome flaws of subjective and behavioural methods discussed above. Physiological methods are continuous, thus enabling real-time monitoring of the state of participants over longer periods of time, without the necessity to interrupt IMEx to gather samples.

Each method strand (subjective, behavioural and psycho-physiological) possess different strengths and weaknesses with regard to IMEx assessment. We therefore argue that a multi-method approach combining all three strands, with each method compensating the disadvantages of the others, would be the only viable way to assess IMEx in all its facets.

# Standardization Activities

**Section Lead**
Christian Timmerer

**Authors**
Kjell Brunnström, Jesús Gutiérrez, Dragan Kukolj, Dragorad Milovanovic, Antonio Pinheiro, Alex Raake, Peter Schelkens, & Christian Timmerer

Interoperability is a major concern for any kind of application and service including those providing immersive media experience. In this section, we provide an overview of standardization activities. According to Timmerer[21], the standardization activities can be grouped into three clusters (bottom up):

1. data representation and formats providing basic tools to be adopted directly by others,

2. guidelines, system standards, and APIs typically providing so-called system specifications including end-to-end aspects, and

3. QoE addressing the perceived quality as experienced by the end users of such applications and services.

*Data representations and formats:* *JPEG*[22] and *MPEG*[23] provide basic tools enabling IMEx within *JPEG Pleno* and *MPEG-I standards*, respectively. The former targets coding tools for omni-directional, depth-enhanced, point cloud, light field, and holographic imaging modalities. The latter comprises omni-directional Media Format (OMAF), Versatile Video Coding (VVC), Video- and Geometry-based Point Cloud Coding, and immersive audio and video coding. *IEEE P.2048.1-12*[24] is a comprehensive set of standards for VR and AR. MPEG-V provides an architecture and specifies associated information representations to enable the interoperability between virtual worlds and with the real world. IEEE P1918.1[25] defines a framework for the Tactile Internet and also ITU-T provides a related report[26].

*Guidelines, system standards, and APIs:* *3GPP*[27] defines codec extensions for VR streaming typically based on MPEG standards. *VRIF*[28] and *DASH-IF*[29] define guidelines based on MPEG's OMAF and DASH/CMAF standards, respectively. *Khronos*[30] specifies OpenXR, which is an API for XR (VR/AR/MR) applications. *WebXR*[31] standardizes APIs to provide access to input and output capabilities commonly associated with XR hardware. *IDEA* defines interoperable interfaces and exchange formats to support the end-to-end conveyance of immersive volumetric and/or light field media. Finally, *CTA*[32], *ETSI*[33], and *SVA*[34] established working groups in the area of IMEx providing guidelines on top of others.

*Quality of Experience (QoE):* QUALINET[35], VQEG[36], ITU-T[37], and IEEE P3333.1.3[38] work on datasets enabling reproducible research, QoE (influence) factors, and subjective/objective quality metrics/assessments in the area of IMEx.

---

21  Timmerer, C. (2017). Immersive media delivery: Overview of ongoing standardization activities. IEEE Communications Standards Magazine, 1(4), 71-74.

22  https://jpeg.org/

23  https://mpeg.chiariglione.org/

24  https://standards.ieee.org/project/2048_1.html

25  https://standards.ieee.org/project/1918_1.html

26  ITU-T, I. T. U. (2014). The Tactile Internet. ITU-T Technology Watch Report.

27  https://www.3gpp.org/

28  https://www.vr-if.org/

29  https://dashif.org/

30  https://www.khronos.org/

31  https://www.w3.org/TR/webxr/

32  https://cta.tech/

33  https://www.etsi.org/

34  https://www.streamingvideoalliance.org/

35  http://www.qualinet.eu/

36  https://www.vqeg.org/

37  https://www.itu.int/

38  https://standards.ieee.org/project/3333_1_3.html

# Final Remarks

The Immersive Media Experience (IMEx) can be seen as an important tool which defines the success of existing and future immersive media applications and services. This white paper provides a toolbox for definitions of IMEx including its Quality of Experience, application areas, influencing factors, and assessment methods. It serves the purpose to enable clarity and guidance for researches and practitioners in both academia and industry working on topics related to IMEx and potentially beyond. Hence, we believe it provides a valuable asset for those working in this field of IMEx towards a better future.

# Acknowledgments

We would like to thank QUALINET, the European Network on Quality of Experience in Multimedia Systems and Services for providing a framework that allows the creation of assets like this white paper and beyond. We also would like to thank the Video Quality Experts Group (VQEG), specifically the Immersive Media Group (IMG) for their valuable feedback during the public community review phase. Finally, we would like to thank the following individuals for their invaluable comments during the community review phase (in alphabetical order): Teresa Chambel, Femi Adeyemi-Ejeye, Marnix van Gisbergen, Patrick Seeling, and Wenhui Zhang.